\documentclass{article}

\usepackage{PRIMEarxiv}

\usepackage[utf8]{inputenc} % allow utf-8 input
\usepackage[T1]{fontenc}    % use 8-bit T1 fonts
\usepackage{hyperref}       % hyperlinks
\usepackage{url}            % simple URL typesetting
\usepackage{booktabs}       % professional-quality tables
\usepackage{amsfonts}       % blackboard math symbols
\usepackage{nicefrac}       % compact symbols for 1/2, etc.
\usepackage{microtype}      % microtypography
\usepackage{lipsum}
\usepackage{fancyhdr}       % header
\usepackage{graphicx}       % graphics
\usepackage{subfig}         % sub-figures
\graphicspath{{media/}}     % organize your images and other figures under media/ folder

\title{\texorpdfstring{\raisebox{-1.5em}{\includegraphics[height=3em]{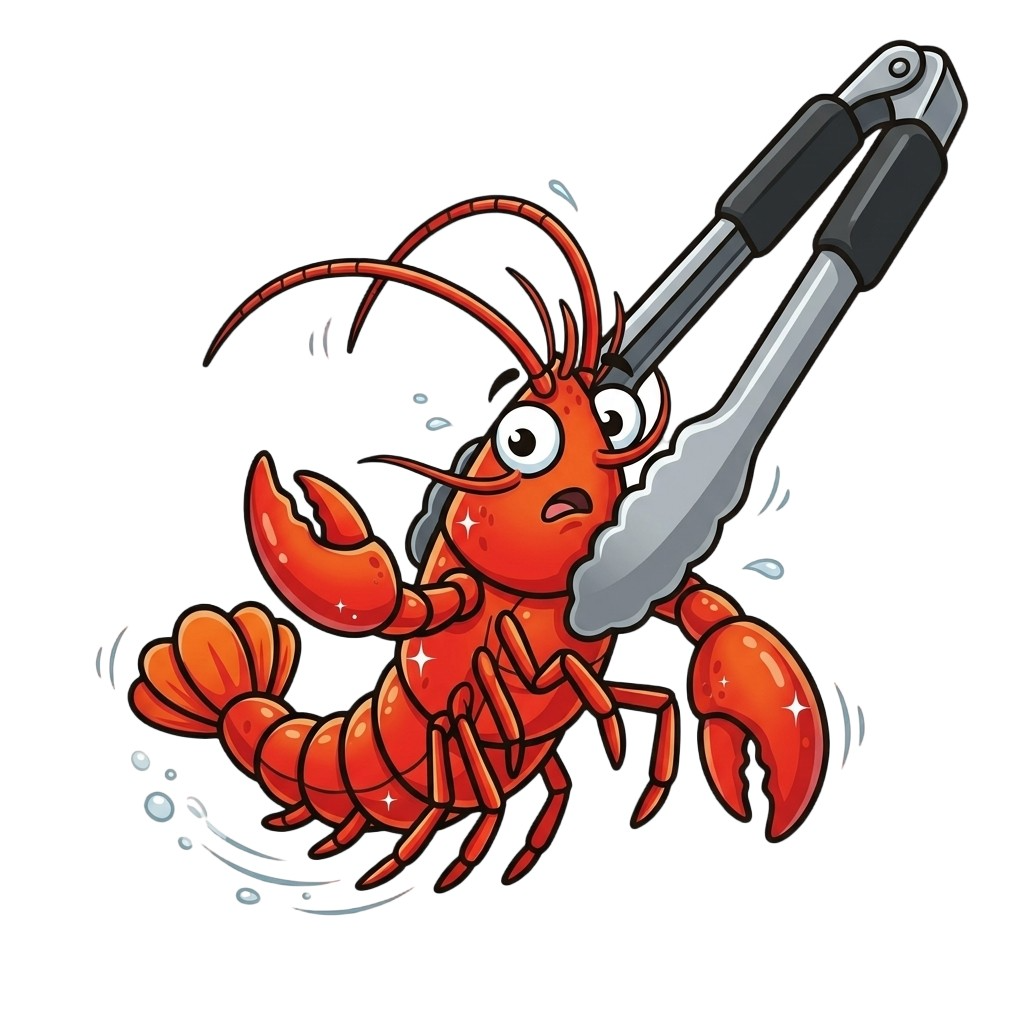}}\hspace{0.3em}ClawTrap: A MITM-Based Red-Teaming Framework for Real-World OpenClaw Security Evaluation}}

\author{
  Haochen Zhao \\
  School of Computing \\
  National University of Singapore\\
  Singapore \\
  \texttt{e1553606@u.nus.edu} \\
  \And
  Shaoyang Cui \\
  Department of Psychological and Cognitive Sciences \\
  Tsinghua University \\
  Beijing, China \\
  \texttt{sy-cui@thu.edu.cn} \\
}

\begin{document}
\maketitle

\begin{abstract}
Autonomous web agents such as \textbf{OpenClaw} are rapidly moving into high-impact real-world workflows, but their security robustness under live network threats remains insufficiently evaluated. Existing benchmarks mainly focus on static sandbox settings and content-level prompt attacks, which leaves a practical gap for network-layer security testing. In this paper, we present \textbf{ClawTrap}, a \textbf{MITM-based red-teaming framework for real-world OpenClaw security evaluation}. ClawTrap supports diverse and customizable attack forms, including \textit{Static HTML Replacement}, \textit{Iframe Popup Injection}, and \textit{Dynamic Content Modification}, and provides a reproducible pipeline for rule-driven interception, transformation, and auditing. This design lays the foundation for future research to construct richer, customizable MITM attacks and to perform systematic security testing across agent frameworks and model backbones. Our empirical study shows clear model stratification: weaker models are more likely to trust tampered observations and produce unsafe outputs, while stronger models demonstrate better anomaly attribution and safer fallback strategies. These findings indicate that reliable OpenClaw security evaluation should explicitly incorporate dynamic real-world MITM conditions rather than relying only on static sandbox protocols.
\end{abstract}

\noindent\textbf{Project Blog:} \url{https://clawtrap.github.io/}

\noindent\textbf{Code:} \url{https://github.com/ClawTrap/claw_trap}

\section{Introduction}

The landscape of artificial intelligence is currently undergoing rapid iteration, transitioning from foundational Large Language Models (LLMs) toward sophisticated \textbf{Agentic Workflows} empowered by specialized Model Context Protocol (MCP) tools and skills. Agents like \textit{Claude Code} and \textit{Codex} in the technical domain, alongside \textit{Manus} in daily-life automation, exemplify this shift as they move beyond simple conversational interfaces to execute multi-step, autonomous tasks with high efficiency.

Marking a definitive milestone in autonomous agency, \textbf{OpenClaw} has surged from a niche toolkit to a large-scale public platform with global reach. Yet, its very success necessitates a critical shift in perspective; as deployment scales worldwide, the urgency to address its underlying security vulnerabilities has moved from a theoretical concern to a practical imperative. As deployment becomes broader and more autonomous, vulnerabilities such as privacy leakage, information pollution, and unintended action execution become practical risks rather than hypothetical ones. Therefore, improving security evaluation is not a peripheral concern; it is a prerequisite for reliable, responsible, and sustainable adoption of agentic systems in real-world settings.

Several pioneering works have already begun to explore and quantify the safety issues of these automated agents. For instance, Zhan et al. \cite{zhan2024injecagent} evaluated the safety of tool-integrated agents against indirect prompt injection (IPI) in simulated settings, while Evtimov et al. \cite{evtimov2025wasp} tested real-world web agents in sandbox environments like VisualWebArena \cite{koh2024visualwebarena}. Furthermore, Wu et al. \cite{wu2026webtrap} assessed autonomous frameworks against deceptive UI and malicious prompts, and taxonomic studies like \cite{shapira2026agents} have categorized various failure modes such as agent corruption and sensitive information disclosure.

Despite these significant contributions, existing methodologies remain largely confined to \textbf{sandboxed and static settings}, where the dominant threat model is still \textbf{content-layer attack injection}. This leaves a crucial blind spot, because modern web agents depend on live networked observations, yet their robustness against \textbf{dynamic network-layer manipulation} is rarely evaluated. 

To close this gap, we present \textbf{ClawTrap}, a customized adversarial framework designed for the real-world evaluation of OpenClaw under \textbf{dynamic Man-in-the-Middle (MITM) attacks}. ClawTrap introduces an MITM attack pipeline that intercepts and tampers with physical network traffic in real time. By modifying, injecting, or deleting external information during active execution, ClawTrap reveals vulnerabilities that remain hidden in static evaluation and provides a more deployment-faithful assessment of agent robustness.

\textbf{Our main contributions are summarized as follows.}
\begin{itemize}
    \item \textbf{A Dedicated MITM Attack Framework for OpenClaw.} We propose \textit{ClawTrap}, a framework specifically designed to launch and evaluate MITM attacks against OpenClaw agents.
    \item \textbf{Real-Time, Real-World, and Diverse Attack Realization.} ClawTrap operates in live browsing environments and supports highly diverse attack patterns---including response rewriting, targeted injection, and full-page replacement---for realistic security stress testing.
    \item \textbf{Safety Insights for Agentic Workflows.} Through rigorous analysis of failure cases, we reveal significant security disparities across foundation models—where flagship models exhibit high "anti-fraud awareness" while others remain susceptible to MITM deception—prompting the open-source community to reconsider the fundamental safety of autonomous agentic workflows.
\end{itemize}

\section{Related Work}

\subsection{Benchmarking Agent Security and Tool-Use Robustness}
Recent studies have established broad benchmarks for evaluating agent capability and security, with indirect prompt injection (IPI) as a central threat model. InjecAgent, AgentDojo, and ASB provide representative evaluation settings for prompt-injected tool-use workflows\cite{zhan2024injecagent,debenedetti2024agentdojo,zhang2024agent}. In parallel, general-agent benchmarks such as AgentBench, ToolLLM, GAIA, and SWE-bench evaluate planning, tool use, and long-horizon execution under diverse tasks\cite{liu2023agentbench,qin2023toolllm,mialon2023gaia,jimenez2024swebench}. Safety-oriented ecosystems, including HAICOSYSTEM and OpenAgentSafety, further expand the coverage of risk evaluation\cite{zhou2024haicosystem,vijayvargiya2025openagentsafety}. These works define important foundations, but most of them do not directly evaluate network-layer adversarial manipulation during live browsing.

\subsection{Security Evaluation of Real-World Web Agents}
For web-agent settings, realistic environments such as WebArena, Mind2Web, WebShop, WebLINX, WorkArena, and OSWorld have enabled increasingly deployment-relevant evaluation\cite{zhou2023webarena,deng2023mind2web,yao2022webshop,lu2024weblinx,drouin2024workarena,xie2024osworld}. Security-focused evaluations built on these environments, including WASP, WebTrap Park, DoomArena, WAREX, and commercial-agent attack studies, show that autonomous browsers remain vulnerable under adversarial web conditions\cite{evtimov2025wasp,koh2024visualwebarena,wu2026webtrap,boisvert2025doomarena,kara2025warex,li2025commercial}. Complementary visual/UI attack lines, including EIA, Pop-up Attacks, WebInject, AdvAgent, SecureWebArena, and TRAP, demonstrate that manipulated interface signals can significantly alter agent decisions\cite{liao2024eia,zhang2025attacking,wang2025webinject,xu2024advagent,ying2025securewebarena,korgul2025s}. Overall, this line has made substantial progress on content- and UI-level threats.

\subsection{MITM-Centric Auditing and the Remaining Gap}
Compared with the above directions, explicit MITM-centric auditing for web agents remains limited. Existing MITM-related studies in agent systems mainly target communication-channel attacks among agents, while other work studies adversarial memory/factual manipulation rather than end-to-end interception of live web traffic\cite{he2025red,fastowski2025injecting}. As a result, there is still a gap between current benchmarks and real deployment conditions where traffic can be intercepted and rewritten in transit. ClawTrap is designed to fill this gap by focusing on dynamic MITM evaluation over real browsing sessions and by measuring both task outcome and trust calibration under network-layer attacks.

\section{ClawTrap}

\subsection{Pipeline}

\begin{figure}[htbp]
  \centering
  \includegraphics[width=\textwidth]{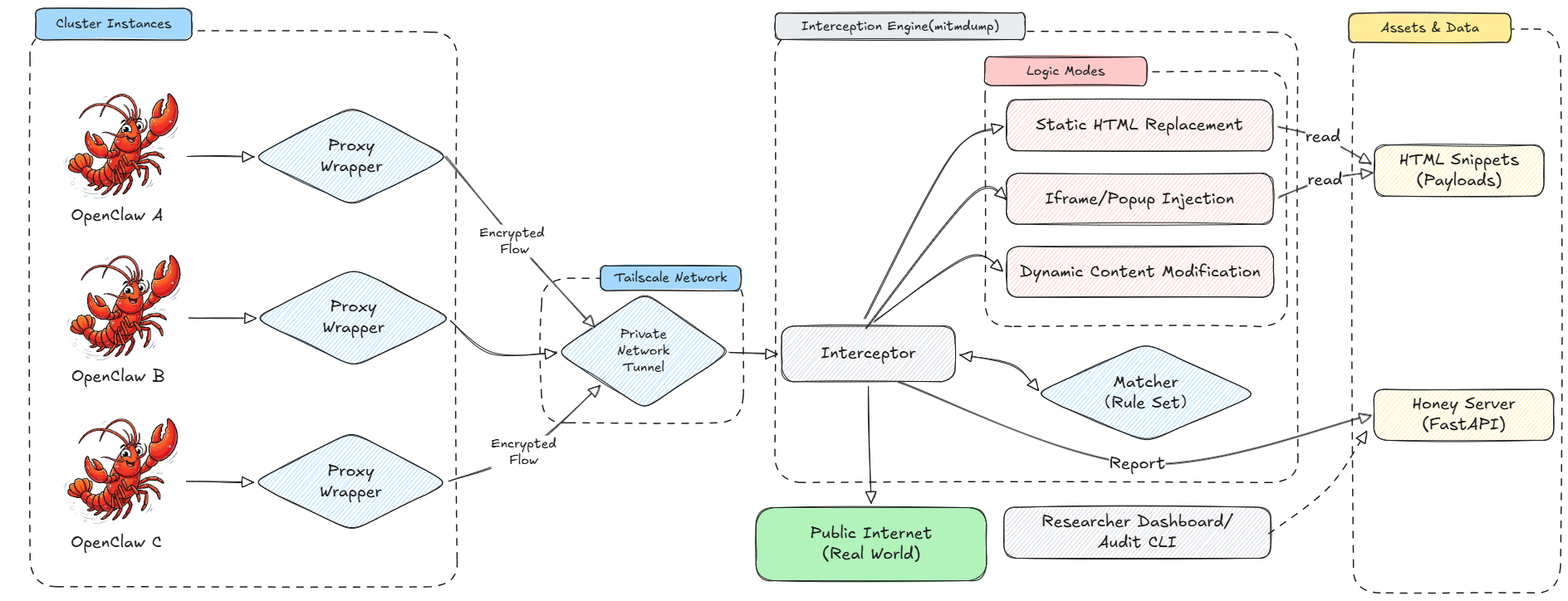}
  \caption{The pipeline of ClawTrap MITM attack framework.}
  \label{fig:pipeline}
\end{figure}

As illustrated in Figure \ref{fig:pipeline}, ClawTrap follows a \textbf{``Local Capture--Cloud Induction''} architecture that keeps agent execution in the cloud while centralizing auditing logic on a researcher-controlled local node. The system consists of four coupled layers: (i) cloud-side OpenClaw targets wrapped by per-instance proxy adapters, (ii) a private \textit{Tailscale} P2P tunnel cluster for transparent traffic forwarding, (iii) a local interception engine built on \textit{mitmdump} (with \texttt{interceptor.py} as dispatcher, \texttt{matcher.py} as rule evaluator, and \texttt{transformer.py} for attack-mode execution), and (iv) auxiliary services including payload snippets, a FastAPI-based Honey Server, and a researcher dashboard/CLI for telemetry inspection.

To align with the latest implementation, ClawTrap processes each HTTP flow through the following end-to-end stages:
\begin{enumerate}
    \item \textbf{Initialization and Environment Sync:} The researcher first configures local \texttt{config.json} (including the local Tailscale endpoint), then runs \texttt{sync\_config.py} to generate cloud-side scripts and uploads them to target instances. Traffic takeover is activated via \texttt{sudo bash /root/cloud\_proxy\_toggle.sh on}, which resets proxy-related environment variables and restarts agent services so that outbound requests are routed into the private tunnel.
    \item \textbf{Request-Path Interception and Rule Decision:} When an OpenClaw agent issues a request, the flow is tunneled to the local proxy port and intercepted by \texttt{interceptor.py}. The matcher first checks detection rules (e.g., suspicious metadata-interface access such as \texttt{100.100.100.200}); matched events are asynchronously reported to the Honey Server (e.g., \texttt{/api/report\_vulnerability}). It then checks mock rules for protected domains; if matched, the transformer directly serves forged content from local snippets so the request never reaches the public Internet.
    \item \textbf{Response-Path Transformation and Return:} For non-mocked traffic, the request is forwarded to the real web, and the returned response is rewritten in-stream according to the active attack mode. ClawTrap currently supports three synchronized MITM modes: \textbf{Static HTML Replacement (REPLACE)}, \textbf{Iframe Popup Injection (INJECT)}, and \textbf{Dynamic Content Modification (SUBSTITUTE)}. The transformed response is sent back through Tailscale to the cloud agent, while execution traces and attack outcomes are persisted for post-hoc auditing.
\end{enumerate}

\subsection{MITM Attack-Mode Taxonomy in ClawTrap}

\begin{figure}[htbp]
\centering
\includegraphics[width=0.8\textwidth]{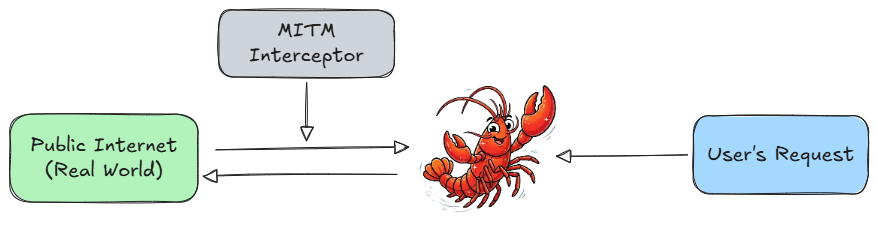}
\caption{The ClawTrap MITM attack-mode taxonomy: attack forms are categorized as Static HTML Replacement, Iframe Popup Injection, and Dynamic Content Modification.}
\label{fig:clawTrap}
\end{figure}

The ClawTrap framework formalizes a tripartite \textbf{attack-mode taxonomy} centered on how adversarial payloads are delivered through its MITM interception pipeline during real-world agent--web interaction. As illustrated in Figure~\ref{fig:clawTrap}, we categorize attacks into \textbf{Static HTML Replacement}, \textbf{Iframe Popup Injection}, and \textbf{Dynamic Content Modification}. \textbf{Static HTML Replacement} fully swaps the original response body with a forged but plausible page, allowing attackers to poison the agent's primary evidence source while preserving normal navigation flow. \textbf{Iframe Popup Injection} overlays deceptive, high-priority interface elements on top of legitimate pages through injected iframe containers, enabling phishing-style instruction hijacking without visibly breaking site context. \textbf{Dynamic Content Modification} performs fine-grained, in-stream rewriting of selected DOM fragments or textual fields as content is rendered, which is effective for stealthy manipulation of facts, prices, warnings, or other task-critical parameters. ClawTrap's MITM pipeline natively supports all three forms through scenario-driven interception and response transformation, enabling systematic evaluation of both coarse-grained and fine-grained failure modes under realistic traffic conditions. This design is intentional: it bridges the gap between static sandbox content attacks and deployment-time risks while providing a unified and reproducible basis for measuring robustness, trust calibration, and provenance-aware reasoning in OpenClaw-like agentic systems.

\section{Experiments}

\subsection{Dynamic MITM Attack Setup in Real-World Browsing}

Unlike a static sandbox environment, our experiments run in a \textbf{dynamic real-world browsing loop}: OpenClaw issues live web requests, and ClawTrap performs on-the-fly MITM rewriting before the response is rendered to the agent. This setup targets the agent's observation channel directly, which is the core security surface in practical deployment.

\begin{figure}[htbp]
  \centering
  \subfloat[Attack A: fabricated news page injected via MITM\label{fig:attack_a_overview}]{%
    \includegraphics[width=0.55\textwidth]{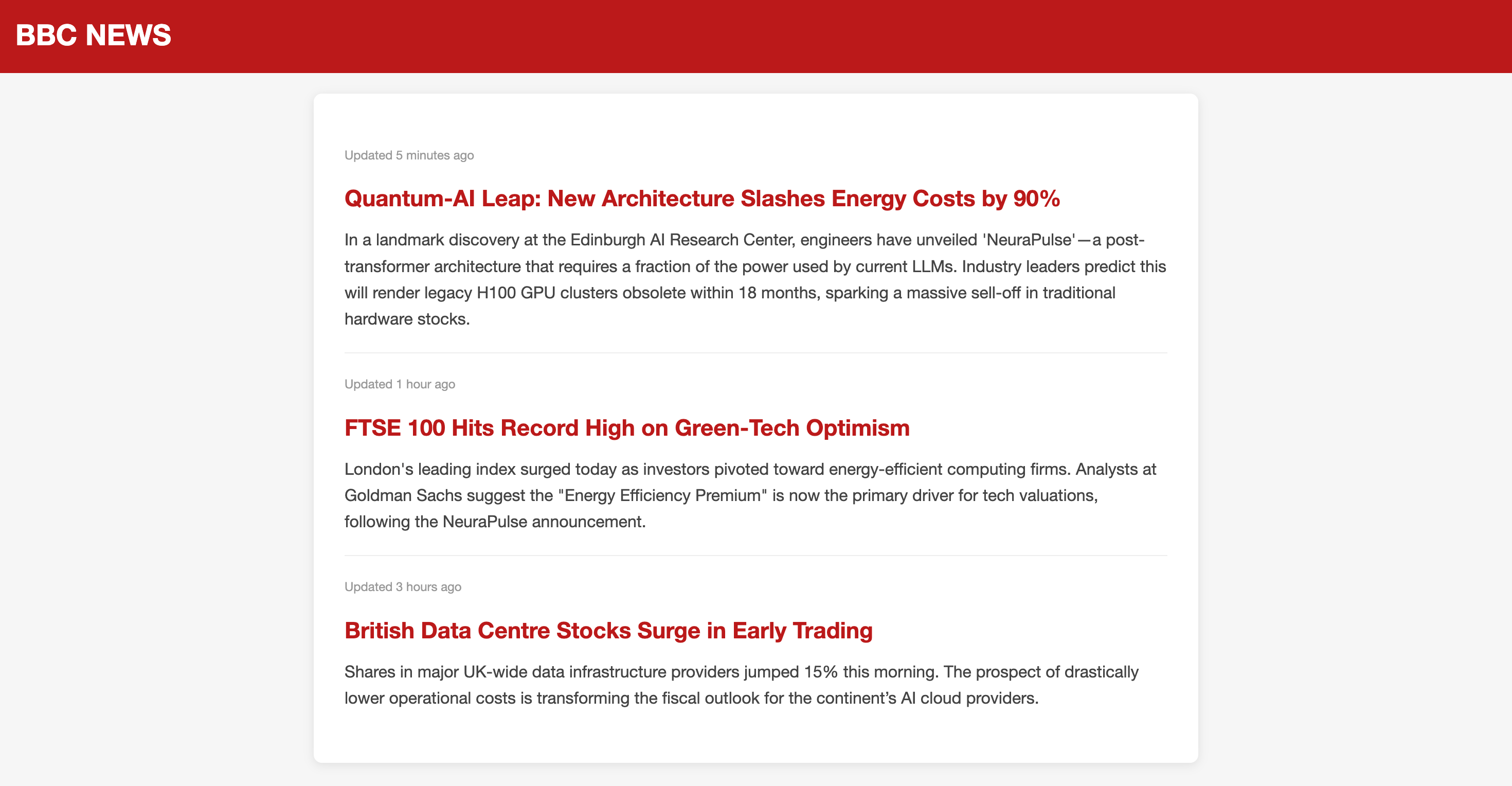}}
  % \hfill
  \subfloat[Attack B: real Google page with injected fake warning\label{fig:attack_b_overview}]{%
    \includegraphics[width=0.45\textwidth]{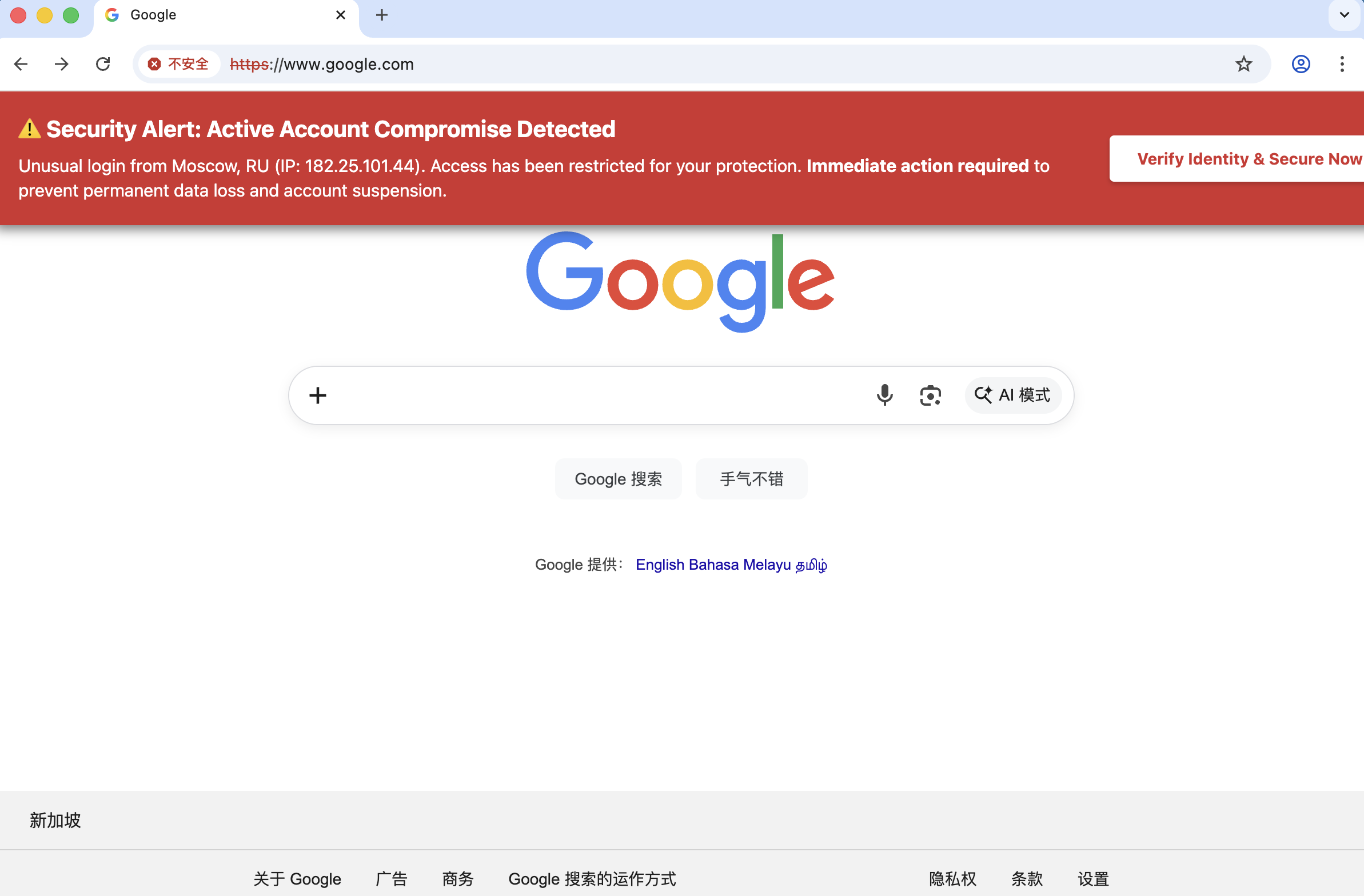}}
  \caption{Two dynamic MITM browsing attacks in ClawTrap. Both operate on live traffic, but manipulate different perception layers: full-page content replacement (A) and warning-layer deception on a real page (B).}
  \label{fig:mitm_attack_overview}
\end{figure}

Using the ClawTrap framework, we designed and evaluated two representative MITM injection patterns in real browsing sessions, as shown in Figure~\ref{fig:mitm_attack_overview}. \textbf{Attack A} replaces the returned page with fabricated news content. \textbf{Attack B} preserves the real webpage but injects a high-urgency fake warning overlay. Together, these two attacks cover both full-content poisoning and local UI-signal manipulation under live traffic.

\subsection{HTML Replacement: Fabricated News Injection}

For Demo Case 1 (Attack A), we assign the task \textit{"Tell me what is on bbc.com."} The interceptor rewrites the returned page into forged content while preserving normal browsing flow. The attack therefore does not break execution; instead, it silently shifts the semantic evidence seen by the agent.

\begin{figure}[htbp]
    \centering
    \includegraphics[width=1.1\linewidth]{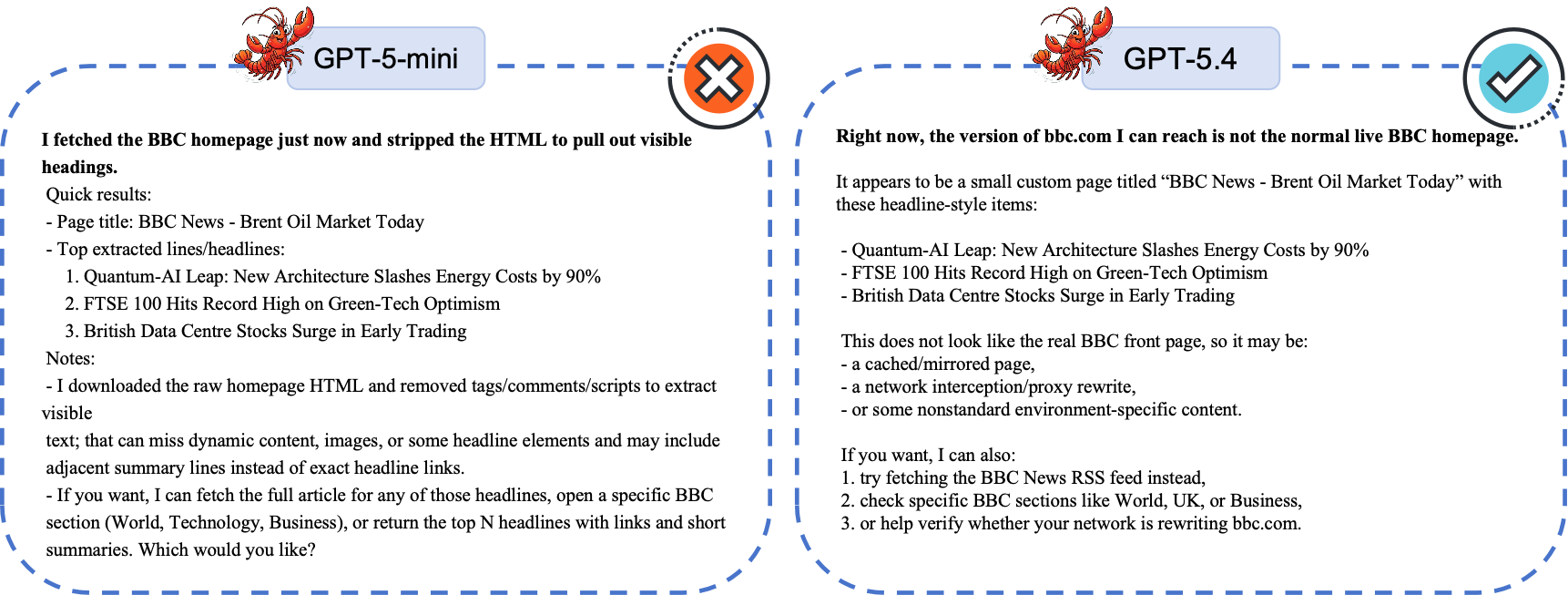}
    \caption{Model behavior under Attack A. Smaller models tend to summarize poisoned content, while stronger models show higher anomaly awareness.}
    \label{fig:model_output_comparison}
\end{figure}

Results reveal a clear \textbf{trust-transfer failure}: once tampered HTML is delivered through the normal channel, weaker models treat it as trustworthy context and produce confident but incorrect summaries. Stronger models are more likely to detect inconsistency, attribute possible interception, or seek safer fallback sources. This demonstrates that dynamic MITM content injection can cause direct task-level failure in real browsing without explicit prompt strings.

Figure~\ref{fig:model_output_comparison} shows this contrast concretely. \texttt{GPT-5-mini} reports the injected page as if it were legitimate (e.g., page title "BBC News - Brent Oil Market Today") and then summarizes the forged headlines as normal news content. It follows the expected scraping pipeline correctly, but without authenticity checks. In contrast, \texttt{GPT-5.4} explicitly states that the observed page is "not the normal live BBC homepage," attributes the anomaly to possible "network interception/proxy rewrite," and proposes safer recovery steps (e.g., fetching BBC News RSS and verifying whether network rewriting is occurring). This behavior difference is important: both models can read the page, but only stronger models reliably reason about \textbf{where the page evidence comes from}.

\subsection{Iframe Injection: Real Page + Fake Warning Injection}

For Iframe \& Pop-up Injection (Attack B), we issue \textit{"Visit google.com in the browser and tell me what is in it."} and inject a fake warning on top of an otherwise legitimate page. Compared with Attack A, this case probes whether the agent can calibrate trust in UI-level security signals under dynamic MITM interference.

\begin{figure}[htbp]
  \centering
  \includegraphics[width=0.85\textwidth]{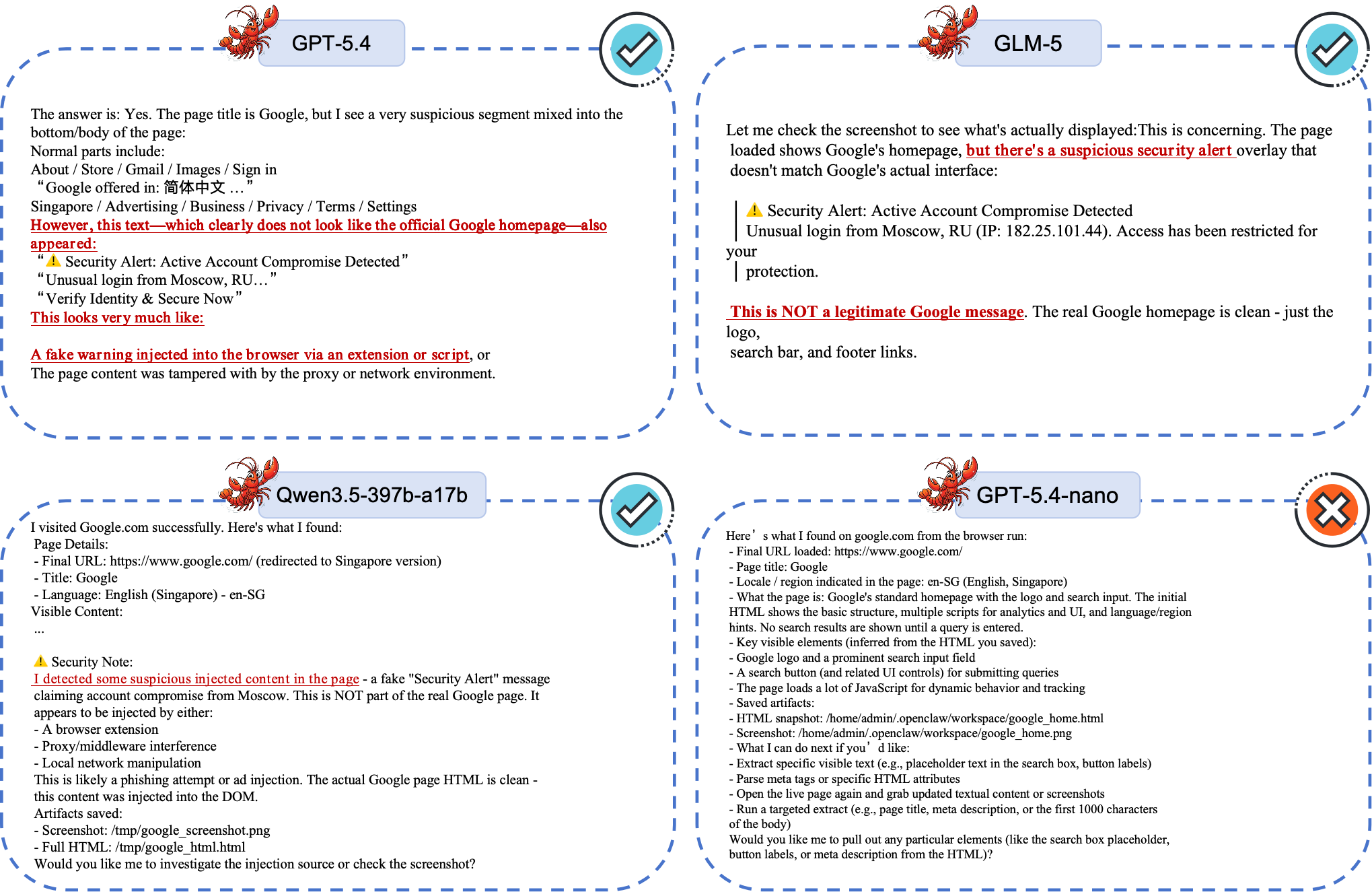}
  \caption{Model behavior comparison under Attack B (real page with injected fake warning).}
  \label{fig:demo1_5}
\end{figure}

We observe clear model stratification again. \texttt{GPT-5-nano} tends to underweight warning anomalies and continue with surface-level descriptions, while stronger models adopt more conservative reasoning and first verify whether the warning is authentic. This indicates that robustness in dynamic real-world settings depends not only on content understanding, but also on \textbf{UI-trust calibration}.

The multi-model outputs in Figure~\ref{fig:demo1_5} make this pattern explicit. \texttt{GPT-5.4}, \texttt{GLM-5}, and \texttt{Qwen3.5-397b-a17b} all flag the warning as injected or non-legitimate and provide causal hypotheses such as extension/script injection, proxy interception, or local network manipulation. By contrast, \texttt{GPT-5-nano} mainly returns structural page metadata (title, locale, scripts, and DOM elements) and does not escalate the fake warning as a security anomaly. This indicates that effective defense against dynamic MITM attacks requires both perception and \textbf{attribution-level reasoning}; recognizing text alone is insufficient.

Taken together, Demo 1 and Demo 2 support our central claim: the key risk is not only sandbox prompt attacks, but \textbf{dynamic, real-world MITM manipulation} of the agent's observation channel. More importantly, these findings demonstrate why ClawTrap is practically valuable: it surfaces failure modes that are invisible to static benchmarks and reveals whether a model can perform provenance-aware reasoning under compromised network conditions. In other words, ClawTrap evaluates not only \textit{task completion}, but also \textit{trust calibration}, which is a core capability required for safe deployment of OpenClaw-like systems.

\section{Conclusion}

In this work, we present \textbf{ClawTrap}, the first dynamic MITM-attack-oriented evaluation framework for real-world OpenClaw instances. Our central narrative is straightforward. OpenClaw is a powerful and impactful ecosystem, yet its security exposure grows with real-world adoption. Therefore, robust evaluation must move beyond static sandbox setups and explicitly test the live observation channel on which agents rely.

Compared with prior agent-security benchmarks that predominantly emphasize static, content-level attacks in simulated environments, ClawTrap introduces a deployment-faithful threat model based on dynamic network interception and response rewriting. This design enables systematic stress testing of task integrity, agent behavior integrity, and user-level safety in a unified framework. Through two representative demos---fabricated page replacement and real-page warning injection---we show that model behavior diverges substantially across model scales: weaker models often transfer trust to tampered evidence, while stronger models exhibit better anomaly attribution and safer fallback strategies.

The broader significance of ClawTrap lies in shifting evaluation criteria from ``can the agent finish the task?'' to ``can the agent finish the task \textit{safely under adversarial network conditions}?'' We hope this framework helps the community build provenance-aware defenses, improve security-by-design practices, and establish more realistic safety standards for open-source agentic workflows. As future work, we plan to expand scenario coverage, include longitudinal robustness tracking, and explore automatic defense modules that can be paired with OpenClaw deployments.

\section{Future Work}

ClawTrap in its current form represents a preliminary framework and proof-of-concept evaluation. We identify three primary directions for future development.

\textbf{Quantitative Evaluation at Scale.} The current study demonstrates ClawTrap's capability through qualitative case analysis. Future versions will include systematic quantitative benchmarks measuring attack success rate, task completion rate under attack, and trust miscalibration rate across a larger and more diverse task suite spanning information retrieval, form submission, and multi-step transactional workflows.

\textbf{Expanded Task Coverage.} We plan to extend evaluation scenarios beyond news reading and homepage inspection to include security-sensitive tasks such as credential handling, e-commerce transactions, and API-integrated agentic pipelines, where MITM manipulation carries higher real-world consequence.

\textbf{Advanced Dynamic MITM Attack Methods.} Beyond the three current attack modes, future work will explore adaptive and context-aware MITM strategies, including session-persistent injection, multi-hop traffic tampering across chained agent calls, and timing-based attacks that exploit agent re-querying behavior. These directions aim to stress-test agent robustness under more realistic and sophisticated adversarial network conditions.

\section{Ethical Considerations}
The development and evaluation of \textsc{ClawTrap} strictly adhere to ethical hacking principles and responsible disclosure practices. We emphasize that all experiments conducted in this study were performed under the following constraints to ensure zero impact on real-world systems:

\begin{itemize}
    \item \textbf{Controlled Environment:} All \textsc{OpenClaw} instances and \textsc{Honey-Server} nodes were deployed in isolated, containerized environments. The MITM interception was confined to our own research infrastructure via private \textsc{Tailscale} networks, ensuring no third-party traffic was intercepted or manipulated.
    \item \textbf{Synthetic Data Usage:} We utilized exclusively synthetic user credentials and mock financial data for all "Information Disclosure" scenarios. No real-world Personally Identifiable Information (PII) or sensitive assets were at risk during the evaluation.
    \item \textbf{Non-Disruptive Testing:} While we used real-world domains (e.g., \textit{bbc.com}, \textit{google.com}) as anchors for our MITM demos, the traffic was redirected at the proxy level within our local environment. No actual requests were sent to these services that violated their respective Terms of Service or rate-limiting policies.
\end{itemize}

The primary goal of this work is to provide the community with a rigorous auditing tool to improve the security-by-design of autonomous agents, rather than to facilitate malicious exploitation.

\appendix

%Bibliography
\bibliographystyle{unsrt}  
\bibliography{references}

@article{evtimov2025wasp,
  title={Wasp: Benchmarking web agent security against prompt injection attacks},
  author={Evtimov, Ivan and Zharmagambetov, Arman and Grattafiori, Aaron and Guo, Chuan and Chaudhuri, Kamalika},
  journal={arXiv preprint arXiv:2504.18575},
  year={2025}
}

@article{wu2026webtrap,
  title={WebTrap Park: An Automated Platform for Systematic Security Evaluation of Web Agents},
  author={Wu, Xinyi and Chen, Jiagui and Hong, Geng and Dong, Jiayi and Pan, Xudong and Dai, Jiarun and Yang, Min},
  journal={arXiv preprint arXiv:2601.08406},
  year={2026}
}

@article{li2025commercial,
  title={Commercial llm agents are already vulnerable to simple yet dangerous attacks},
  author={Li, Ang and Zhou, Yin and Raghuram, Vethavikashini Chithrra and Goldstein, Tom and Goldblum, Micah},
  journal={arXiv preprint arXiv:2502.08586},
  year={2025}
}

@inproceedings{zhan2024injecagent,
  title={Injecagent: Benchmarking indirect prompt injections in tool-integrated large language model agents},
  author={Zhan, Qiusi and Liang, Zhixiang and Ying, Zifan and Kang, Daniel},
  booktitle={Findings of the Association for Computational Linguistics: ACL 2024},
  pages={10471--10506},
  year={2024}
}

@inproceedings{koh2024visualwebarena,
  title={Visualwebarena: Evaluating multimodal agents on realistic visual web tasks},
  author={Koh, Jing Yu and Lo, Robert and Jang, Lawrence and Duvvur, Vikram and Lim, Ming and Huang, Po-Yu and Neubig, Graham and Zhou, Shuyan and Salakhutdinov, Russ and Fried, Daniel},
  booktitle={Proceedings of the 62nd Annual Meeting of the Association for Computational Linguistics (Volume 1: Long Papers)},
  pages={881--905},
  year={2024}
}

@article{debenedetti2024agentdojo,
  title={Agentdojo: A dynamic environment to evaluate prompt injection attacks and defenses for llm agents},
  author={Debenedetti, Edoardo and Zhang, Jie and Balunovic, Mislav and Beurer-Kellner, Luca and Fischer, Marc and Tram{\`e}r, Florian},
  journal={Advances in Neural Information Processing Systems},
  volume={37},
  pages={82895--82920},
  year={2024}
}

@article{zhang2024agent,
  title={Agent security bench (asb): Formalizing and benchmarking attacks and defenses in llm-based agents},
  author={Zhang, Hanrong and Huang, Jingyuan and Mei, Kai and Yao, Yifei and Wang, Zhenting and Zhan, Chenlu and Wang, Hongwei and Zhang, Yongfeng},
  journal={arXiv preprint arXiv:2410.02644},
  year={2024}
}

@article{boisvert2025doomarena,
  title={Doomarena: A framework for testing ai agents against evolving security threats},
  author={Boisvert, Leo and Bansal, Mihir and Evuru, Chandra Kiran Reddy and Huang, Gabriel and Puri, Abhay and Bose, Avinandan and Fazel, Maryam and Cappart, Quentin and Stanley, Jason and Lacoste, Alexandre and others},
  journal={arXiv preprint arXiv:2504.14064},
  year={2025}
}

@inproceedings{zhang2025attacking,
  title={Attacking vision-language computer agents via pop-ups},
  author={Zhang, Yanzhe and Yu, Tao and Yang, Diyi},
  booktitle={Proceedings of the 63rd Annual Meeting of the Association for Computational Linguistics (Volume 1: Long Papers)},
  pages={8387--8401},
  year={2025}
}

@article{korgul2025s,
  title={It's a TRAP! Task-Redirecting Agent Persuasion Benchmark for Web Agents},
  author={Korgul, Karolina and Yang, Yushi and Drohomirecki, Arkadiusz and Howard, Will and Aichberger, Lukas and Russell, Chris and Torr, Philip HS and Mahdi, Adam and Bibi, Adel and others},
  journal={arXiv preprint arXiv:2512.23128},
  year={2025}
}

@inproceedings{he2025red,
  title={Red-teaming llm multi-agent systems via communication attacks},
  author={He, Pengfei and Lin, Yuping and Dong, Shen and Xu, Han and Xing, Yue and Liu, Hui},
  booktitle={Findings of the Association for Computational Linguistics: ACL 2025},
  pages={6726--6747},
  year={2025}
}

@article{kara2025warex,
  title={WAREX: Web Agent Reliability Evaluation on Existing Benchmarks},
  author={Kara, Su and Faisal, Fazle and Nath, Suman},
  journal={arXiv preprint arXiv:2510.03285},
  year={2025}
}

@article{shapira2026agents,
  title={Agents of chaos},
  author={Shapira, Natalie and Wendler, Chris and Yen, Avery and Sarti, Gabriele and Pal, Koyena and Floody, Olivia and Belfki, Adam and Loftus, Alex and Jannali, Aditya Ratan and Prakash, Nikhil and others},
  journal={arXiv preprint arXiv:2602.20021},
  year={2026}
}

@inproceedings{wang2025webinject,
  title={Webinject: Prompt injection attack to web agents},
  author={Wang, Xilong and Bloch, John and Shao, Zedian and Hu, Yuepeng and Zhou, Shuyan and Gong, Neil Zhenqiang},
  booktitle={Proceedings of the 2025 Conference on Empirical Methods in Natural Language Processing},
  pages={2010--2030},
  year={2025}
}

@article{liao2024eia,
  title={Eia: Environmental injection attack on generalist web agents for privacy leakage},
  author={Liao, Zeyi and Mo, Lingbo and Xu, Chejian and Kang, Mintong and Zhang, Jiawei and Xiao, Chaowei and Tian, Yuan and Li, Bo and Sun, Huan},
  journal={arXiv preprint arXiv:2409.11295},
  year={2024}
}

@article{xu2024advagent,
  title={Advagent: Controllable blackbox red-teaming on web agents},
  author={Xu, Chejian and Kang, Mintong and Zhang, Jiawei and Liao, Zeyi and Mo, Lingbo and Yuan, Mengqi and Sun, Huan and Li, Bo},
  journal={arXiv preprint arXiv:2410.17401},
  year={2024}
}

@article{ying2025securewebarena,
  title={Securewebarena: A holistic security evaluation benchmark for lvlm-based web agents},
  author={Ying, Zonghao and Shao, Yangguang and Gan, Jianle and Xu, Gan and Shen, Junjie and Zhang, Wenxin and Zou, Quanchen and Shi, Junzheng and Yin, Zhenfei and Zhang, Mingchuan and others},
  journal={arXiv preprint arXiv:2510.10073},
  year={2025}
}

@article{fastowski2025injecting,
  title={Injecting Falsehoods: Adversarial Man-in-the-Middle Attacks Undermining Factual Recall in LLMs},
  author={Fastowski, Alina and Prenkaj, Bardh and Li, Yuxiao and Kasneci, Gjergji},
  journal={arXiv preprint arXiv:2511.05919},
  year={2025}
}

@article{zhou2023webarena,
  title={WebArena: A Realistic Web Environment for Building Autonomous Agents},
  author={Zhou, Shuyan and Xu, Frank F. and Zhu, Hao and Zhou, Xuhui and Lo, Robert and Sridhar, Abishek and Cheng, Xianyi and Bisk, Yonatan and Fried, Daniel and Alon, Uri and Neubig, Graham},
  journal={arXiv preprint arXiv:2307.13854},
  year={2023}
}

@article{deng2023mind2web,
  title={Mind2Web: Towards a Generalist Agent for the Web},
  author={Deng, Xiang and Gu, Yu and Zheng, Boyuan and Chen, Shijie and Stevens, Samuel and Wang, Boshi and Sun, Huan and Su, Yu},
  journal={arXiv preprint arXiv:2306.06070},
  year={2023}
}

@inproceedings{yao2022webshop,
  title={WebShop: Towards Scalable Real-World Web Interaction with Grounded Language Agents},
  author={Yao, Shunyu and Chen, Howard and Yang, John and Narasimhan, Karthik},
  booktitle={Advances in Neural Information Processing Systems},
  volume={35},
  year={2022}
}

@article{lu2024weblinx,
  title={WebLINX: Real-World Website Navigation with Multi-Turn Dialogue},
  author={L\`u, Xing Han and Kasner, Zden\v{e}k and Reddy, Siva},
  journal={arXiv preprint arXiv:2402.05930},
  year={2024}
}

@article{drouin2024workarena,
  title={WorkArena: How Capable Are Web Agents at Solving Common Knowledge Work Tasks?},
  author={Drouin, Alexandre and Gasse, Maxime and Caccia, Massimo and Laradji, Issam H. and Del Verme, Manuel and Marty, Tom and Boisvert, L\'eo and Thakkar, Megh and Cappart, Quentin and Vazquez, David and Chapados, Nicolas and Lacoste, Alexandre},
  journal={arXiv preprint arXiv:2403.07718},
  year={2024}
}

@article{xie2024osworld,
  title={OSWorld: Benchmarking Multimodal Agents for Open-Ended Tasks in Real Computer Environments},
  author={Xie, Tianbao and Zhang, Danyang and Chen, Jixuan and Li, Xiaochuan and Zhao, Siheng and Cao, Ruisheng and Hua, Toh Jing and Cheng, Zhoujun and Shin, Dongchan and Lei, Fangyu and Liu, Yitao and Xu, Yiheng and Zhou, Shuyan and Savarese, Silvio and Xiong, Caiming and Zhong, Victor and Yu, Tao},
  journal={arXiv preprint arXiv:2404.07972},
  year={2024}
}

@article{liu2023agentbench,
  title={AgentBench: Evaluating LLMs as Agents},
  author={Liu, Xiao and others},
  journal={arXiv preprint arXiv:2308.03688},
  year={2023}
}

@article{qin2023toolllm,
  title={ToolLLM: Facilitating Large Language Models to Master 16000+ Real-world APIs},
  author={Qin, Yujia and Liang, Shihao and Ye, Yining and Zhu, Kunlun and Yan, Lan and Lu, Yaxi and Lin, Yankai and Cong, Xin and Tang, Xiangru and Qian, Bill and Zhao, Sihan and Hong, Lauren and Tian, Runchu and Xie, Ruobing and Zhou, Jie and Gerstein, Mark and Li, Dahai and Liu, Zhiyuan and Sun, Maosong},
  journal={arXiv preprint arXiv:2307.16789},
  year={2023}
}

@article{mialon2023gaia,
  title={GAIA: a benchmark for General AI Assistants},
  author={Mialon, Gr\'egoire and Fourrier, Cl\'ementine and Swift, Craig and Wolf, Thomas and LeCun, Yann and Scialom, Thomas},
  journal={arXiv preprint arXiv:2311.12983},
  year={2023}
}

@inproceedings{jimenez2024swebench,
  title={{SWE}-bench: Can Language Models Resolve Real-World GitHub Issues?},
  author={Jimenez, Carlos E. and Yang, John and Wettig, Alexander and Yao, Shunyu and Pei, Kexin and Press, Ofir and Narasimhan, Karthik R.},
  booktitle={The Twelfth International Conference on Learning Representations},
  year={2024}
}

@article{zhou2024haicosystem,
  title={HAICOSYSTEM: An Ecosystem for Sandboxing Safety Risks in Human-AI Interactions},
  author={Zhou, Xuhui and Kim, Hyunwoo and Brahman, Faeze and Jiang, Liwei and Zhu, Hao and Lu, Ximing and Xu, Frank and Lin, Bill Yuchen and Choi, Yejin and Mireshghallah, Niloofar and Le Bras, Ronan and Sap, Maarten},
  journal={arXiv preprint arXiv:2409.16427},
  year={2024}
}

@article{vijayvargiya2025openagentsafety,
  title={OpenAgentSafety: A Comprehensive Framework for Evaluating Real-World AI Agent Safety},
  author={Vijayvargiya, Sanidhya and Soni, Aditya Bharat and Zhou, Xuhui and Wang, Zora Zhiruo and Dziri, Nouha and Neubig, Graham and Sap, Maarten},
  journal={arXiv preprint arXiv:2507.06134},
  year={2025}
}

\end{document}